\documentclass[a4paper]{spie}  %>>> use this instead for A4 paper
\usepackage[]{graphicx, url}
\title{Image reconstruction for observations with a high dynamic range: LINC-NIRVANA simulations of a stellar jet} 

\author{Andrea {La Camera}\supit{a}, Simone Antoniucci\supit{b}, Mario Bertero\supit{a},
 Patrizia Boccacci\supit{a}, Dario Lorenzetti\supit{b}, Brunella Nisini\supit{b}
\skiplinehalf
\supit{a}Dipartimento interscuola di Informatica, Bioingegneria, Robotica e Ingegneria dei Sistemi (DIBRIS), Universit\`a di Genova, 
    Via all'Opera Pia 13, 16145, Genova, Italy;\\
\supit{b}INAF - Osservatorio Astronomico di Roma, Via di Frascati 33, 00040, 
    Monteporzio Catone, Italy
}

\authorinfo{Further author information: (Send correspondence to A.L.C., S.A.)\\A.L.C.: E-mail: \url{andrea.lacamera@unige.it}, Telephone: +39 010 353 6981\\  S.A.: E-mail: \url{simone.antoniucci@oa-roma.inaf.it}, Telephone: +39 06 94286 487}
\begin{document} 
  \maketitle 

%%%%%%%%%%%%%%%%%%%%%%%%%%%%%%%%%%%%%%%%%%%%%%%%%%%%%%%%%%%%% 
\begin{abstract}
We report the results of a simulation and reconstruction of observations of a 
young stellar object (YSO) jet with the LINC-NIRVANA (LN) interferometric instrument, which will be mounted on the Large Binocular Telescope (LBT). 
This simulation has been performed in order to investigate the ability of 
observing the weak diffuse jet line emission against the strong IR stellar 
continuum through narrow band images in the H and K atmospheric windows. 
In general, this simulation provides clues on the image quality that could be 
achieved in observations with a high dynamic range. 
In these cases, standard deconvolution methods, such as Richardson-Lucy, do not 
provide satisfactory results: we therefore propose here a new method of image 
reconstruction. It consists in considering the image to be reconstructed  as 
the sum of two terms: one corresponding to the star (whose position is assumed 
to be known) and the other to the jet. 
A regularization term is introduced for this second component and the 
reconstruction is obtained with an iterative method alternating between the two 
components. 
An analysis of the results shows that the image quality obtainable with this 
method is significantly improved with respect to standard deconvolution methods, 
reducing the number of artifacts and allowing us to reconstruct the original 
jet intensity distribution with an error smaller than 10\%.
\end{abstract}

%>>>> Include a list of keywords after the abstract 

\keywords{Fizeau interferometry, image restoration, high dynamic range imaging, 
Richardson-Lucy algorithm, LINC-NIRVANA}

%%%%%%%%%%%%%%%%%%%%%%%%%%%%%%%%%%%%%%%%%%%%%%%%%%%%%%%%%%%%%
\section{INTRODUCTION}
\label{sec:intro}  % \label{} allows reference to this section

The Large Binocular Telescope (LBT), currently operating on
Mount Graham in Arizona, consists of two 8.4m mirrors on the same
mount. This peculiar structure is particularly suited to perform Fizeau interferometry and,
to this purpose, the near-infrared image-plane beam combiner
LINC-NIRVANA (LN) is in an advanced stage of realization by an
Italian-German consortium led by the Max Planck Institute for
Astronomy in Heidelberg.

\begin{figure}
\begin{center}
\begin{tabular}{cc}
\fbox{\includegraphics[width=0.4\textwidth]{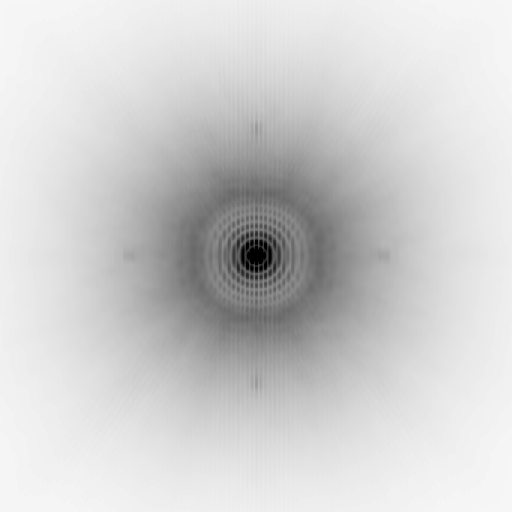}}& 
\fbox{\includegraphics[width=0.4\textwidth]{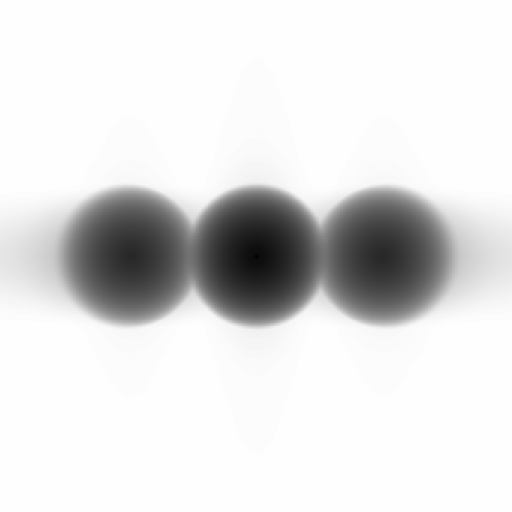}} \\ 
 & \\ 
\end{tabular}
\caption{Simulated PSF of LN with SR=70\% (left panel), and the corresponding 
MTF (right panel). The fringes are orthogonal to the baseline. Both images are 
shown in log scale.}
\label{fig:psf+mtf}
\end{center}
\end{figure}

The performance of LN is expected to be close to the diffraction limit.
Due to the binocular structure of the instrument, the Point Spread 
Function (PSF) may be described as the diffraction limited pattern of
a 8.4m mirror crossed by the fringes due to the interference between the
two apertures, with a maximum baseline of approximately 22.8m.
In Fig. \ref{fig:psf+mtf} we give an example of PSF with SR=0.7, obtained with 
the software LOST{\cite{arcidiacono}}, together with its Modulus Transfer 
Function (MTF). Since
the resolution of a given LN image is anisotropic, several images
with different orientations of the baseline must be acquired and
then processed (and combined) to obtain a final image with high
resolution in all directions. For a discussion of the imaging
problem for LINC-NIRVANA we refer to the review article by
Bertero et al.{\cite{bertero}}.

High angular resolution observations obtainable with an
interferometer like LN are essential in the study of young jets, providing
information on the magneto-hydrodynamical process that is at the
origin and collimation of the jet. It is thus crucial to investigate the
morphology of the jet in the close surroundings of the star,
as the relevant physical processes are at work in the
first 10 AU from the star, which correspond to about 70 mas (i.e. 14 pixels 
in our simulation) for objects in the nearest star-forming clouds at a 
distance of $\sim$150 pc. For this reason, it is of paramount importance to 
understand the goodness of the image reconstruction of the jet achievable 
by LN in the region around the star.

To this aim, in Ciliegi et al.{\cite{ciliegi}}, LN-images of a YSO model 
were generated using the software package AIRY-LN{\cite{desidera}} and they 
were reconstructed using a multiple-image deconvolution method derived from
the well known Richardson-Lucy (RL){\cite{richardson,lucy}} iterative
algorithm. However, the reconstruction of the jet was unsatisfactory in the
region close to the emitting star, which is the most important for studying
the physical process responsible for the launching and collimation of the jet.
The main reason of this result is that standard deconvolution methods
are unable to reconstruct an astronomical target consisting of a
bright point-wise component (typically a star) superimposed to a weaker,
diffuse component (in our case, the emitted jet). A bibliographical
analysis of the methods proposed for solving this image reconstruction
problem is provided by Giovannelli \& Coulais{\cite{giovannelli}}. In
addition these authors propose, in the case of radio interferometry,
an approach which consists in considering the object to be
reconstructed as a superposition of two components, a point source and
an extended source, and in introducing different regularization terms
for the two components within a least-square approach to the problem.
This is equivalent to assume an additive Gaussian noise.

In this paper we develop a similar idea but in the framework of Poisson
noise where the standard image reconstruction method is RL. In the
case of multiple image deconvolution, the problem arising in LINC-NIRVANA
imaging, extensions of RL must be considered{\cite{bertero}}. The method
we propose will be denoted as multi-component Richardson-Lucy (MC-RL)
method.

%%%%%%%%%%%%%%%%%%%%%%%%%%%%%%%%%%%%%%%%%%%%%%%%%%%%%%%%%%%%%
\section{THE MC-RL METHOD}
\label{sec:algo}  % \label{} allows reference to this section

As stated in Sect.~\ref{sec:intro} we assume that the object to be 
reconstructed $f = f({\bf n})$, where ${\bf n}=\{n_1,n_2\}$ is a multi-index 
labeling the pixels of the image, is the sum of two terms (or components), 
$f({\bf n})= f_1({\bf n})+f_2({\bf n})$, where $f_1$ corresponds to the
point source and $f_2$ to the extended source. For simplicity we assume
that the source is a single star whose position is known with a pixel
precision, so that $f_1({\bf n})= c~ \delta({\bf n}, {\bf n}_0)$, where 
${\bf n}_0$ is the pixel where the star is located, $c$ is its unknown
emission intensity and $\delta$ denotes the usual delta function. We first 
describe the approach in the case of single-image deconvolution; next we 
extend it to the case of multiple images.  

\subsection{Single-image deconvolution}

As it is well-known, in the case of Poisson noise the negative logarithm
of the likelihood is given, except for constant factors, by the
Kullback-Leibler divergence, also known as Csisz{\'a}r I-divergence{\cite{csiszar}}. Therefore, under the assumptions above, maximum likelihood
(ML) solutions can be obtained by minimizing with respect to $f_1,f_2$ the
following function
\begin{equation}
J_0(f_1,f_2;g) = \sum_{{\bf n} \in S}
\left\{g({\bf n}){\rm ln}\frac{g({\bf n})}{[A(f_1+f_2)]({\bf n})+ 
b({\bf n})} +[A(f_1+f_2)]({\bf n})+b ({\bf n}) -g({\bf n}) \right\}~~,
\label{KL}
\end{equation}
where $b$ is the background emission (assumed to be known) and $A$ is the
imaging matrix, which is given in terms of a space-invariant point spread 
function (PSF) $K$, normalized to unit volume, as follows
\begin{equation}
(Af)({\bf n})=(K*f)({\bf n})~~,~~\sum_{{\bf n} \in S}K({\bf n})=1~~.
\label{immatrix}
\end{equation}
The function $J_0$ is nonnegative, convex and coercive in $f_1,f_2$, as well 
as in their sum $f=f_1+f_2$ and therefore ML solutions exist. However, it is
well-known that they appear as {\it sky-night solutions}{\cite{barrett}}, i.e.
a set of bright spots over a black background. While this result can be
satisfactory for the point source component, it is not for the extended
source component. Therefore we add a regularization term for
this one, a procedure which can be justified in the framework of a
Bayesian approach. In conclusion, the function we intend to minimize has
the following structure
\begin{equation}
J_\mu(f_1,f_2;g)=J_0(f_1,f_2;g)+
\frac{\mu}{2} \sum_{{\bf n} \in S} |f_2({\bf n})|^2~~,
\label{KLreg}
\end{equation}
where we assume a Tikhonov-like regularization and $\mu$ is a regularization
parameter to be estimated. We discuss this point in the next section.

In order to extend RL to the minimization of this function with respect to 
$f_1, f_2$, we must first compute its gradient with respect to these
variables. We use the well-known expression of the gradient of $J_0(f;g)$, 
we denote as $\nabla_1, \nabla_2$ the gradients with respect to $f_1, f_2$,
respectively, and finally, for simplifying the notation, we denote as $f$ 
their sum $f=f_1+f_2$. Using the normalization of the PSF it follows that
\begin{eqnarray}
\nabla_1J_\mu(f_1,f_2;g)={\hat 1} - A^T \frac{g}{Af+b}~~, \\ \nonumber
\nabla_2J_\mu(f_1,f_2;g)={\hat 1} - A^T \frac{g}{Af+b}+ \mu f_2~~,
\label{gradient}
\end{eqnarray}
where $\hat 1$ is the array with all elements equal to 1 and the quotient of 
two arrays is intended pixel by pixel.

If we use, for instance, the {\it split-gradient method} (SGM){\cite{lanteri}}, 
then the iterative method for the minimization of $J_\mu$ is as follows:
given $f_{1}^{(0)}, f_{2}^{(0)}$, for $k=0,1,..$ compute
\begin{eqnarray}
\label{iteration}
f_1^{(k+1)}=f_1^{(k)}A^T\frac{g}{Af^{(k)}+b} \\ \nonumber
f_2^{(k+1)}=\frac{f_2^{(k)}}{{\hat 1}+\mu f_2^{(k)}}A^T\frac{g}{Af^{(k)}+b} 
\\ \nonumber
f^{(k+1)}=f_1^{(k+1)}+f_2^{(k+1)}~~.
\end{eqnarray}
It is easy to check by induction that, if we initialize the algorithm with
$f_1^{(0)}({\bf n})= \delta({\bf n},{\bf n}_0)$ (and $f_2^{(0)}$ arbitrary
but positive) then, at iteration $k$, we have 
$f_{1}^{(k)}({\bf n})=c_k \delta({\bf n},{\bf n}_0)$. 

We do not have a proof of convergence of the previous algorithm, even if we 
have always found convergence in our numerical experiments. The algorithm is
very slow, even slower than the standard RL algorithm. However, since it is
a scaled gradient method, convergence and acceleration can be obtained in
the framework of the so-called {\it scaled gradient projection} (SGP) 
method{\cite{BZZ}}. 

\subsection{Multiple-image deconvolution}

In the case of $p$ images, $g_1,...,g_p$ (we denote as $g$ the set formed by 
these images), if we assume their statistical independence, then 
Eq. (\ref{KL}) is replaced by
\begin{equation}
J_0(f_1,f_2;g) = \sum_{j=1}^{p}\sum_{{\bf n} \in S}
\left\{g_j({\bf n}){\rm ln}\frac{g_j({\bf n})}{[A_j(f_1+f_2)]({\bf n})+
b_j ({\bf n})} +[A_j(f_1+f_2)]({\bf n})+b_j ({\bf n}) -g_j({\bf n}) \right\}~~,
\label{KLmultiple}
\end{equation}
and the gradients of $J_\mu(f_1,f_2;g)$, defined as in Eq. (\ref{KLreg}), 
are given by
\begin{eqnarray}
\nabla_1J_\mu(f_1,f_2;g)=p{\hat 1} - \sum_{j=1}^p A_j^T \frac{g_j}{A_jf+b}~~, 
\\ \nonumber
\nabla_2J_\mu(f_1,f_2;g)=p{\hat 1}-\sum_{j=1}^p A_j^T 
\frac{g_j}{A_jf+b}+ \mu f_2~~.
\label{gradient1}
\end{eqnarray}
Therefore the algorithm (\ref{iteration}) is replaced by: given $f_{1}^{(0)}, f_{2}^{(0)}$, for $k=0,1,...$
compute
\begin{eqnarray}
f_1^{(k+1)}=\frac{1}{p}f_1^{(k)}\sum_{j=1}^p A_j^T\frac{g_j}{A_jf^{(k)}+b} \\ 
\nonumber
f_2^{(k+1)}=\frac{f_2^{(k)}}{p{\hat 1}+\mu f_2^{(k)}}
\sum_{j=1}^p A_j^T\frac{g_j}{A_jf^{(k)}+b} 
\\ \nonumber
f^{(k+1)}=f_1^{(k+1)}+f_2^{(k+1)}~~.
\label{multiiter}
\end{eqnarray}
The remarks concerning the algorithm (\ref{iteration}) apply also to this one.

%%%%%%%%%%%%%%%%%%%%%%%%%%%%%%%%%%%%%%%%%%%%%%%%%%%%%%%%%%%%%
\section{NUMERICAL EXPERIMENTS}
\label{sec:test}  % \label{} allows reference to this section

\subsection{Description of the observation simulation}
The synthetic image for our simulation has been obtained from an optical
image taken with HST of the HH34 jet{\cite{reip}}. Starting from this
image in Ciliegi et al.{\cite{ciliegi}} the pixel scale was changed to 
the 5mas/pixel of LN and the integrated magnitude of the object was 
normalized to 13 mag in the H$_{2}$ band ($\lambda=2.12 \mu m$, 
$\Delta \lambda=0.02 \mu m$); moreover a point source having an H$_{2}$ 
magnitude of 13 mag was added at the position where the HH34 infrared driving 
source is located; finally, the average K-band sky brightness of 13.5 
mag/arcsec$^2$ has been assumed as background emission. 

Four equispaced $1024 \times 1024$ LN images, at 
$0^\circ,~45^\circ,~90^\circ$ and $135^\circ$, have been obtained by 
convolving the object with four PSFs generated through the software 
LOST{\cite{arcidiacono}}. The results are corrupted with Poisson and 
additive Gaussian noise with $\sigma=10 \ e^{-}/$pixel. For each hour angle 
an integration time of 30 minutes has been chosen, for a total integration 
time of 2 hours. In Fig. 2 we show the object and one of the simulated 
LN-images. 

\begin{figure}[!ht]
\begin{center}
\begin{tabular}{cc}
\fbox{\includegraphics[width=0.3\textwidth]{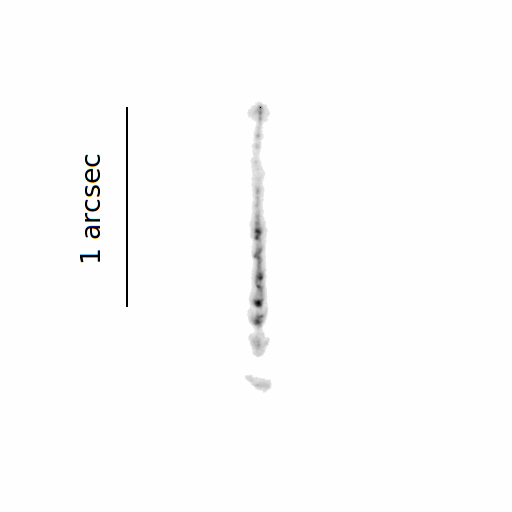}} & 
\fbox{\includegraphics[width=0.3\textwidth]{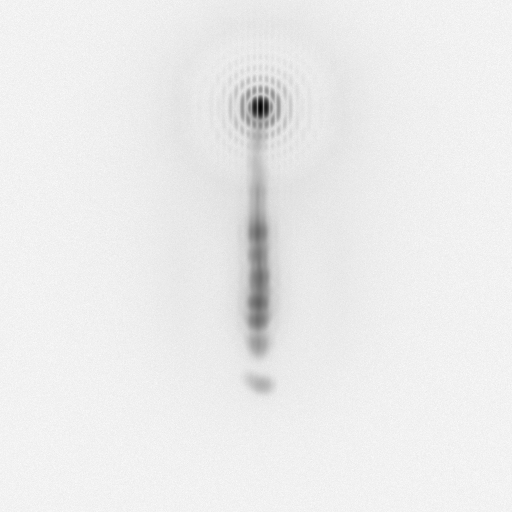}} \\
 & \\ 
\fbox{\includegraphics[width=0.3\textwidth]{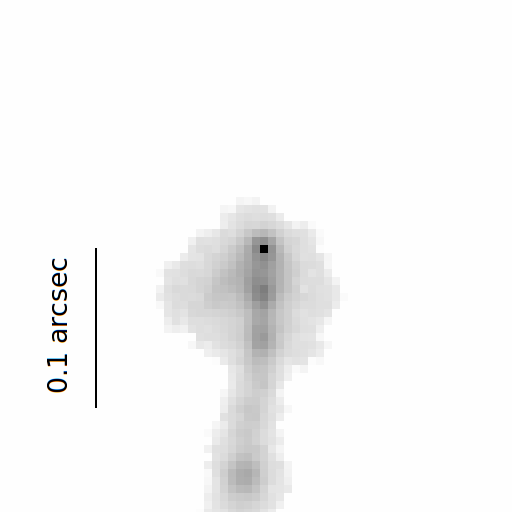}} & 
\fbox{\includegraphics[width=0.3\textwidth]{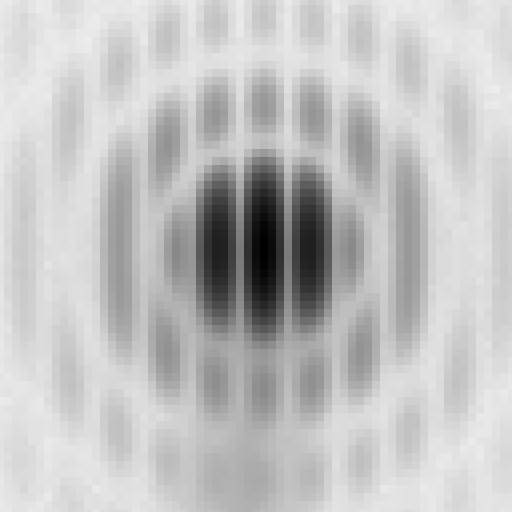}} \\
 & \\ 
\end{tabular}
\caption{The upper panels show the starting image of HH34 used for the 
simulations (left) and the simulated LN image at $0^\circ$ (right). The lower 
panels display a zoomed view of the region surrounding the star (left) and the 
corresponding part in the simulated image (right). The object images 
(left panels) are shown in sqrt scale, whereas the simulated LN observations 
(right panels) are in log scale.}
\label{fig:obj+img}
\end{center}
\end{figure}

\subsection{Image deconvolution}
The image deconvolution has been performed using the MC-RL method described in 
the previous section and, for comparison purposes, the standard Richardson-Lucy 
method for multiple images (RLM). In both cases, for the deconvolution process, 
we utilized the same PSFs used in convolution (\emph{inverse crime}): in such 
a way we obtained the best reconstructed object available from the input data. 

Since the two algorithms are iterative, early stopping of the iterations
is required for obtaining sensible solutions. A stopping rule that can
be used in numerical simulations, since we have the complete knowledge of the 
true object $\bar{f}$, consists in stopping the iteration when the relative 
r.m.s. error (also called \emph{restoration error}), defined by

\begin{equation}
\rho^{(k)} = \frac{||f^{(k)}- \bar{f} ||}{|| \bar{f}||},
\end{equation} 
reaches a minimum value. As pointed out in Sect.~{\ref{sec:intro}}, 
MC-RL is able to separately reconstruct the star and the diffuse part of the 
object, while RLM is not. For this reason, we computed the restoration error in 
two slightly different ways for the two algorithms:
\begin{itemize}
\item RLM: we re-defined the true object $\bar{f}$ by using the image shown in 
the upper-left panel of Fig. \ref{fig:obj+img} (i.e. the star superimposed to 
the jet) setting to 0 the values of the object in a 3$\times$3 region around 
the star position. In the same way we masked the reconstructed image before 
computing the restoration error.
\item MC-RL: we computed the restoration error by using  $\bar{f}=\bar{f_{2}}$ 
(i.e. the diffuse jet without the star) as shown in Fig. 
\ref{fig:reconstructions}.
\end{itemize} 

Moreover, since the astrophysical interest for this kind of object is focused 
on the region close to the star (where formation and collimation of the jet 
occur, see Sect.~\ref{sec:analysis}), we computed two different restoration 
errors: the former (indicated by $\rho'$) by considering the entire jet, the 
latter ($\rho''$) by considering only 24 rows of the image around the star 
position. 
 
\begin{figure}
\begin{center}
\begin{tabular}{cc}
\fbox{\includegraphics[width=0.45\textwidth]{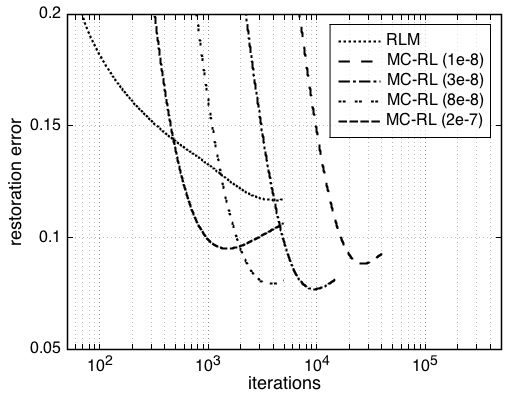}} & 
\fbox{\includegraphics[width=0.45\textwidth]{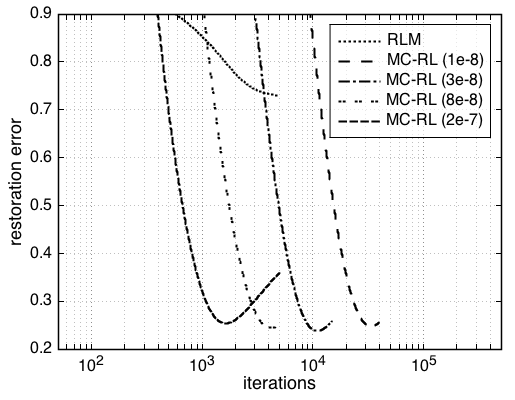}} \\
 & \\ 
\end{tabular}
\caption{The restoration error $\rho'$ of the entire jet (left panel), and 
the restoration error $\rho''$ of the small region close to the star 
(right panel). Both errors are a function of the number of iterations.}
\label{fig:plot_rho}
\end{center}
\end{figure}

As concerns the MC-RL method, the choice of the regularization parameter $\mu$ 
can be performed by minimizing the restoration error defined in the previous 
equation. For the values of $\mu$=$2\cdot10^{-7}$, $8\cdot10^{-8}$, 
$3\cdot10^{-8}$, and $1\cdot10^{-8}$, we plot in Fig. \ref{fig:plot_rho} 
the two restoration errors $\rho'$ and $\rho''$ as functions of the number of
iterations. In addition, we also plot the restoration errors of the 
reconstructed object obtained by RLM. We report the minimum values and the 
number of iterations at which they occur in Tab. \ref{tab:rho}. 

\begin{table}[!ht]
\begin{center}
\begin{tabular}{c|cc}
\hline
Method & $\rho'$ (nb of iterations) & $\rho''$ (nb of iterations) \\
\hline
RLM & 11.67\% (4251) & 72.87\% ($>5000$) \\
MC-RL ($\mu=1\cdot10^{-8}$) & 8.82\% (26614)& 24.80\% (33483) \\
MC-RL ($\mu=3\cdot10^{-8}$) & 7.68\% (9503) & 23.81\% (10972) \\
MC-RL ($\mu=8\cdot10^{-8}$) & 7.91\% (3776) & 24.42\% (4198) \\
MC-RL ($\mu=2\cdot10^{-7}$) & 9.49\% (1496) & 25.34\% (1630) \\
\hline
\multicolumn{3}{c}{}
\end{tabular}
\caption{The minimum values of the restoration errors $\rho'$ and $\rho''$ 
and the corresponding number of iterations. }
\label{tab:rho}
\end{center}
\end{table}

%%%%%%%%%%%%%%%%%%%%%%%%%%%%%%%%%%%%%%%%%%%%%%%%%%%%%%%%%%%%%
\section{DATA ANALYSIS}
\label{sec:analysis}  % \label{} allows reference to this section

\begin{figure}[!ht]
\begin{center}
\begin{tabular}{ccc}
\fbox{\includegraphics[width=0.27\textwidth]{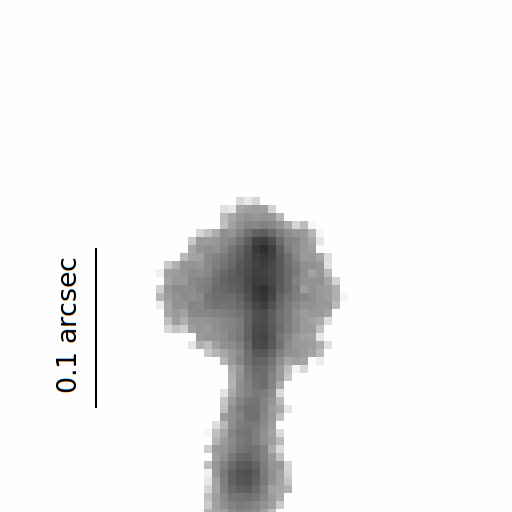}} & 
\fbox{\includegraphics[width=0.27\textwidth]{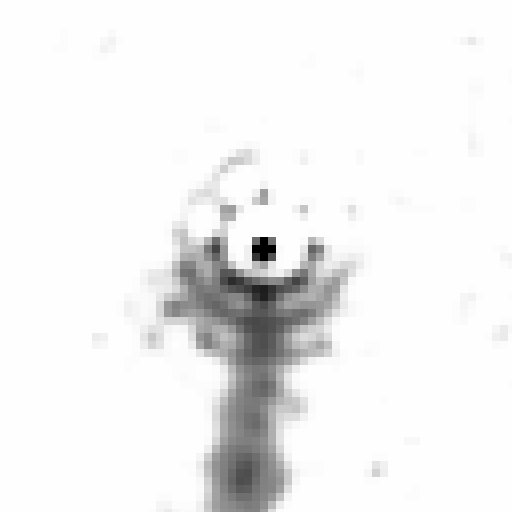}} & 
\fbox{\includegraphics[width=0.27\textwidth]{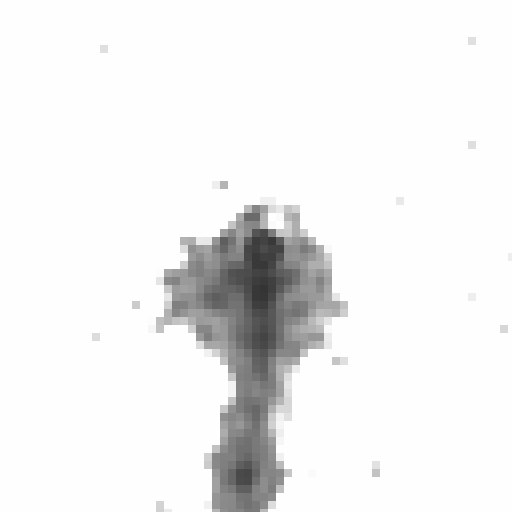}} \\
 {\small True object} & {\small RLM} & {\small MC-RL ($\mu=1\cdot10^{-8}$)} \\
& & \\
\fbox{\includegraphics[width=0.27\textwidth]{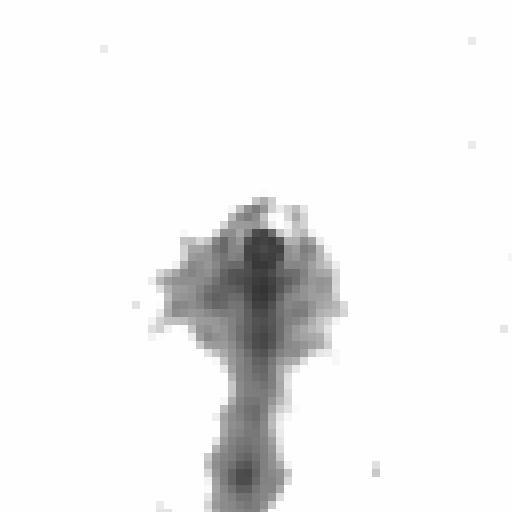}} & 
\fbox{\includegraphics[width=0.27\textwidth]{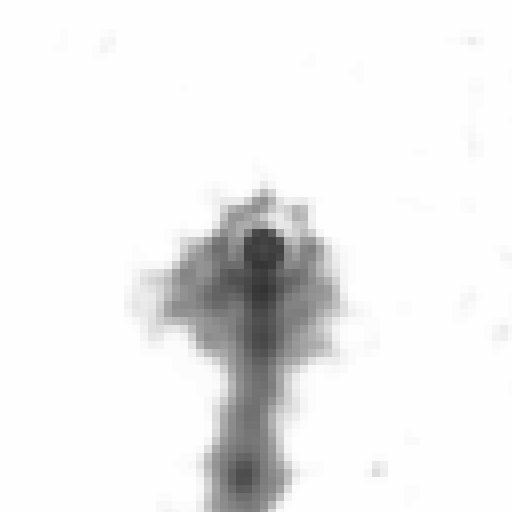}} & 
\fbox{\includegraphics[width=0.27\textwidth]{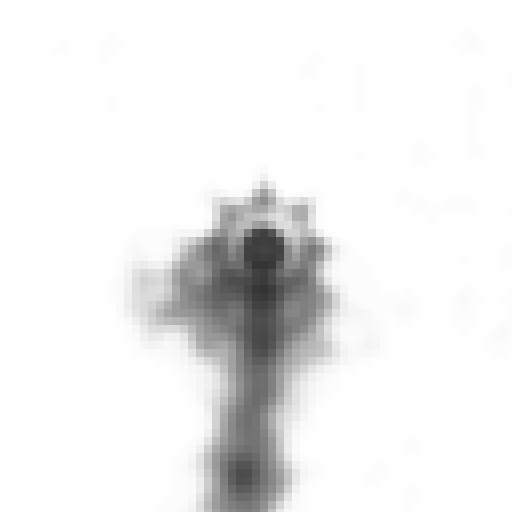}} \\
 {\small MC-RL ($\mu=3\cdot10^{-8}$)} & {\small MC-RL ($\mu=8\cdot10^{-8}$)} 
& {\small MC-RL ($\mu=2\cdot10^{-7}$)} \\
 & & \\
\end{tabular}
\caption{The true object, the reconstructed object obtained with RLM, and the 
reconstructions with MC-RL for different values of the regularization 
parameter $\mu$.}
\label{fig:reconstructions}
\end{center}
\end{figure}

\begin{figure}[!ht]
\begin{center}
\includegraphics[width=0.7\textwidth]{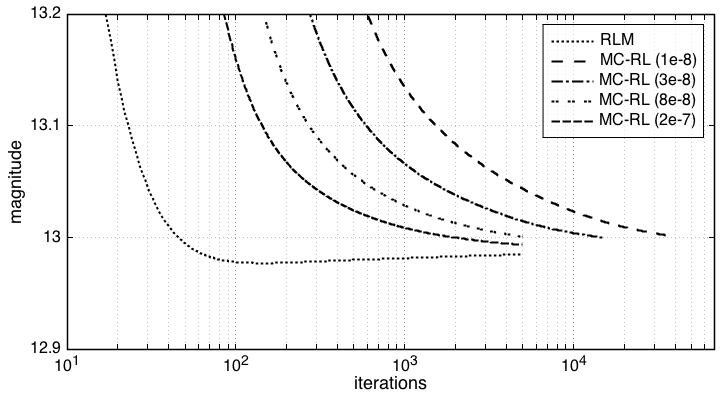}
\caption{The magnitude of the reconstructed star for RLM and MC-RL. All 
reconstructions provide the correct magnitude (m=13) of the star, with errors
smaller than 0.1~mag.}
\label{fig:mag}
\end{center}
\end{figure}

In Fig.~\ref{fig:reconstructions} we show the reconstructed object for the 
RLM case and for the MC-RL cases with different values of $\mu$. More precisely, 
a magnification of the part of the jet close to the star is shown. These 
images correspond to the reconstructions obtained with the number of 
iterations that provides the minimum value of $\rho''$ (see Table~\ref{tab:rho}). 
Moreover, in Fig.~\ref{fig:mag} we show the magnitude of the reconstructed 
star as a function of the number of iterations. The star flux has been computed 
in a 3$\times$3 box centered on its known position. Although there 
are small differences between RLM and MC-RL, both algorithms provide the correct 
value (13 mag) with satisfactory precision. Moreover, RLM reaches this value 
after a smaller number of iterations than the MC-RL method.

However, as remarked in Sect.\ref{sec:intro},  it is very important to 
investigate the morphology of the jet in the close surroundings of the star 
and, for this reason, we need to obtain a good image reconstruction of the 
jet in the region around the star. In this context, the significant difference 
that we find for the restoration error $\rho''$ (see Tab.~\ref{tab:rho}) 
already indicates that the MC-RL method works much better than the standard RLM. 
Of course, since the restoration error provides only an indication of the global 
quality of the reconstructions, a more detailed analysis requires a direct 
comparison of the object morphology in the original and reconstructed images. 
Indeed, a quick look at Fig.~\ref{fig:reconstructions} clearly shows that the 
region around the star is significantly less affected by artifacts when using 
the MC-RL method, confirming the result indicated by $\rho''$.

In order to compare the different methods, we have tried to quantify the 
quality of the reconstructions by taking into account two additional parameters: 
the integrated flux in the direction perpendicular to the jet axis (total 
profile flux) and the jet width. The first one is simply the flux obtained by 
summing up each pixel line of the image, as the jet is oriented along the 
y-axis of the image itself. As for the width computation, we cannot use a 
Gaussian FWHM, since the profile of the jet is in general not Gaussian. Hence, 
we set an arbitrary count threshold and define the width of the jet as the 
number of pixels lying between the first and the last pixel (along the 
considered line) that have a number of counts greater than the threshold.
We note that the width is a key parameter for the scientific analysis, since 
it is directly related to the measurement (to be performed as close as possible 
to the exciting source) of the collimation of the jet. 

\begin{figure}[!ht]
\begin{center}
\begin{tabular}{cc}
\includegraphics[width=0.47\textwidth]{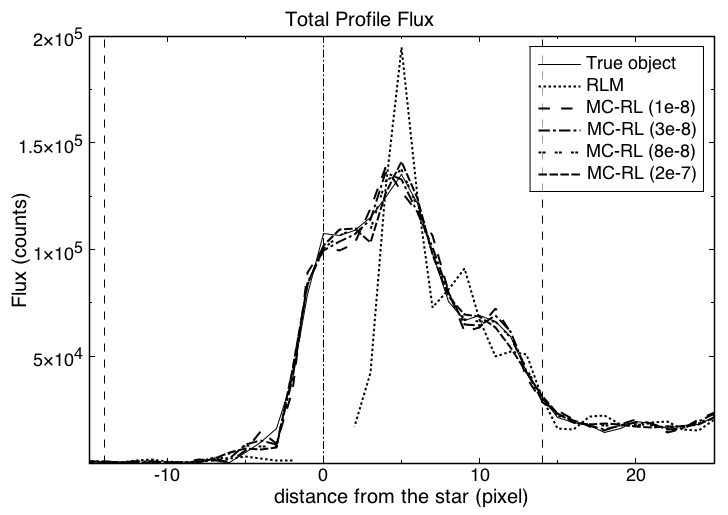} & 
\includegraphics[width=0.47\textwidth]{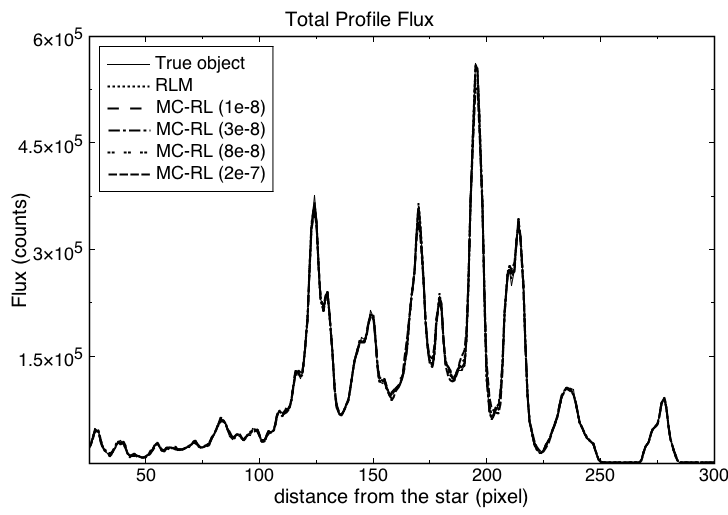} \\
\includegraphics[width=0.47\textwidth]{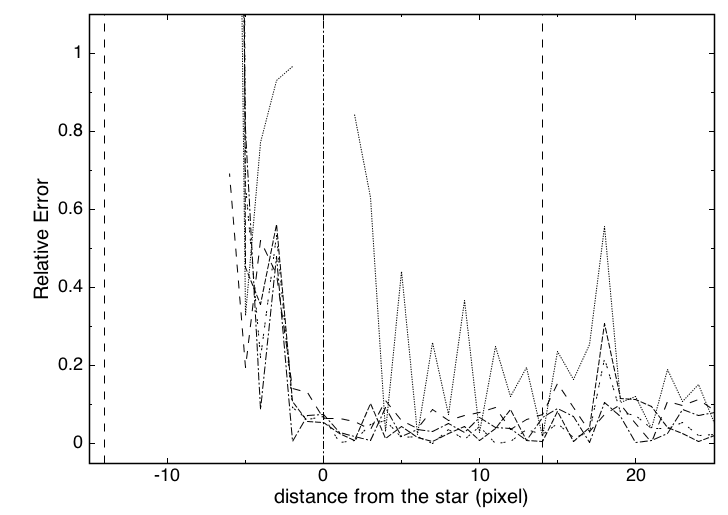} & 
\includegraphics[width=0.47\textwidth]{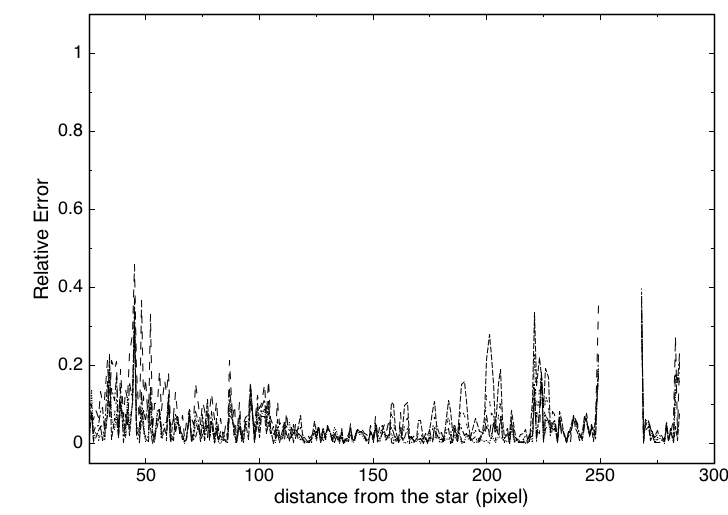} \\
\multicolumn{2}{c}{}
\end{tabular} 
\caption{Comparison of the integrated jet profile flux as measured in the 
original object image (solid thick line) and in the various reconstruction 
methods considered (RLM and MC-RL with different values of $\mu$, see legend) 
as a function of the position along the jet axis. The visualization is split 
in two plots: one for the region around the star (left panels), covering 
$\sim$40 pixels, and the other including the remaining part of the jet 
(right panels). The position of the star is marked by the vertical dashed-dotted 
line, while the dashed vertical lines delimit the typical region of interest 
for the formation and collimation of the jet (within $\sim$10~AU from the star)
in the case of an object at a distance of 150~pc. The differences between the 
parameter values measured in the original image and in each reconstructed 
image are displayed in terms of relative errors in the bottom panels.
In order to compare the jet reconstruction obtained through RLM (containing 
the star) and the results from MC-RL (where the star is not present), we have 
masked the 3 pixel lines around the position of the star in the RLM image.}
\label{fig:analysis_flux}
\end{center}
\end{figure}

\begin{figure}[!ht]
\begin{center}
\begin{tabular}{cc}
\includegraphics[width=0.47\textwidth]{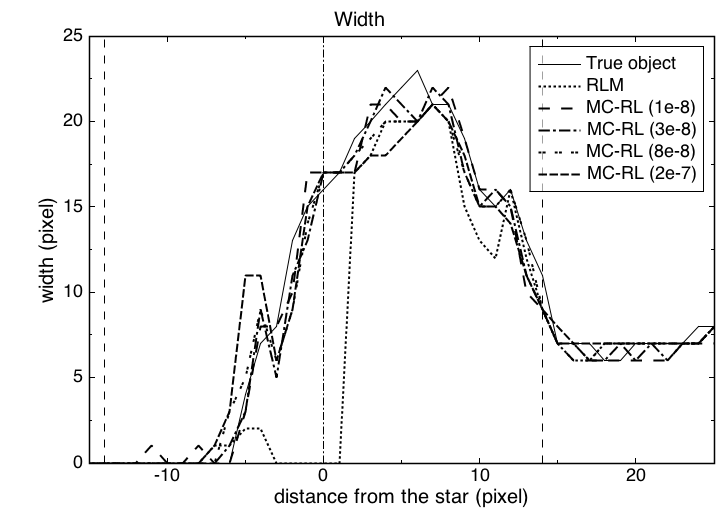} & 
\includegraphics[width=0.47\textwidth]{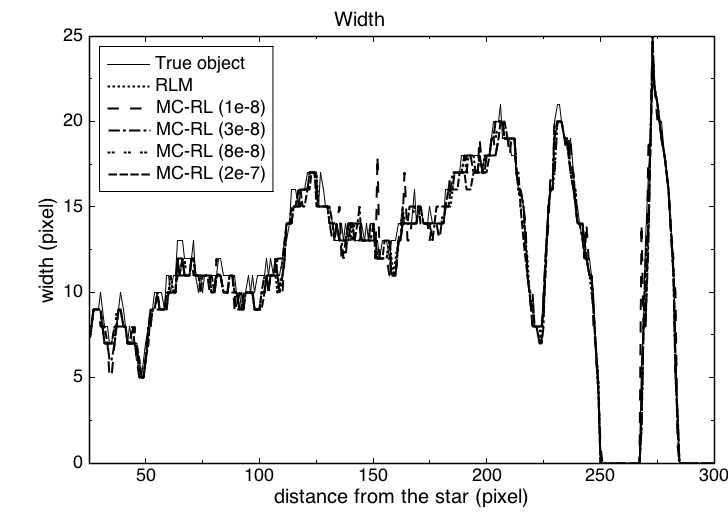} \\ 
\includegraphics[width=0.47\textwidth]{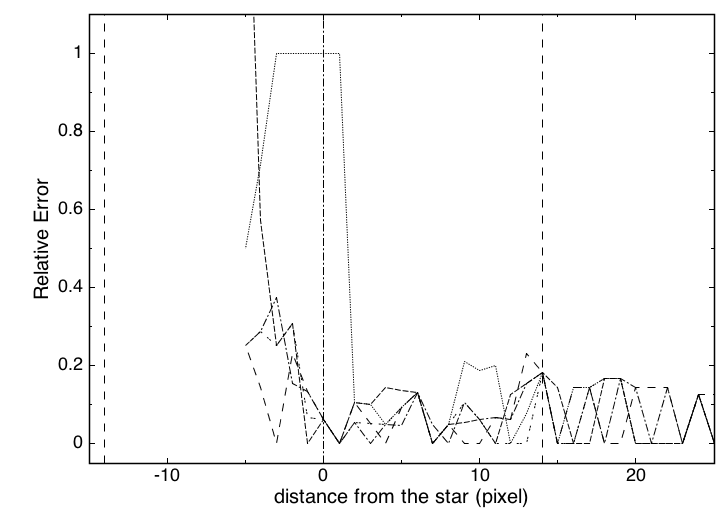} & 
\includegraphics[width=0.47\textwidth]{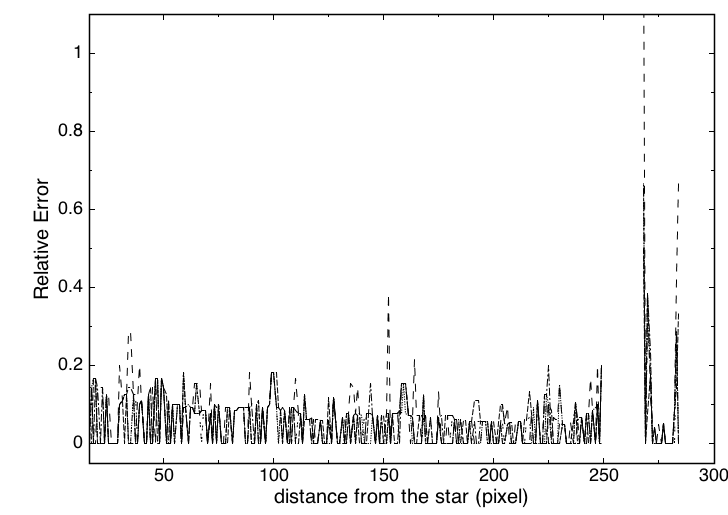} \\
\multicolumn{2}{c}{}
\end{tabular} 
\caption{Comparison of jet width as measured in the original object image 
(solid thick line) and in the various reconstruction methods considered 
(RLM and MC-RL with different values of $\mu$, see legend) as a function of 
the position along the jet axis. The visualization is split in two plots: one 
for the region around the star (left panels), covering $\sim$40 pixels, and the 
other including the remaining part of the jet (right panels). The position of 
the star is marked by the vertical dashed-dotted line, while the dashed 
vertical lines delimit the typical region of interest for the formation and 
collimation of the jet (within $\sim$10~AU from the star) in the case of an 
object at a distance of 150~pc. The differences between the parameter values 
measured in the original image and in each reconstructed image are displayed 
in terms of relative errors in the bottom panels.}
\label{fig:analysis_width}
\end{center}
\end{figure}

We analyze the total profile flux (shown in Fig.~\ref{fig:analysis_flux}) and 
width (Fig.~\ref{fig:analysis_width}) as a function of the position along the jet and directly compare the values measured on the original image to the ones 
obtained for RLM and MC-RL images. In particular, we show the differences with 
respect to the original image in terms of the relative error on the parameter 
values. For each parameter we have used two separate series of plots: one for 
the region around the star, covering $\sim$200~mas (40 pixels), and the other 
including the rest of the jet.

The plots show that in the jet region both RLM and MC-RL methods provide a 
fairly good reconstruction of the image, with relative errors typically below 
10\% for both total profile flux and width. We note that MC-RL with the lowest 
and highest $\mu$ values ($1\cdot10^{-8}$ and $2\cdot10^{-7}$) give somewhat 
larger relative errors for the profile flux (peaks up to $\sim$20\%-30\%), so 
that intermediate $\mu$ values appear to work better.

In the star region the RLM shows instead much higher relative errors for the 
total profile flux (40\%-60\%) because of the artifacts, whereas MC-RL 
reconstructions are characterized by errors that are on average around 10\% only.
As for the width measurement, the best reconstructions are again obtained using 
the MC-RL algorithm with intermediate values of $\mu$ ($\mu=8\cdot10^{-8}$ 
and $\mu=3\cdot10^{-8}$), for which we get errors of about 10\% in proximity 
of the star.

To summarize, our analysis shows that the use of the MC-RL algorithm allows us 
to limit the number and intensity of the artifacts in the region around the star. 
In the case examined, intermediate values of the regularization parameter 
$\mu$ provide the most accurate reconstruction of the original image. This 
shows that the MC-RL method can provide optimal results through a fine-tuning 
of this parameter.

%%%%%%%%%%%%%%%%%%%%%%%%%%%%%%%%%%%%%%%%%%%%%%%%%%%%%%%%%%%%%
\section{CONCLUSIONS}
\label{sec:conclusions}
We have simulated the observations and image reconstructions of a YSO jet with the LBT/LINC-NIRVANA 
interferometric instrument. 
This is a typical case of high dynamic range observations (weak diffuse jet line emission 
superimposed to a strong stellar continuum) in which standard deconvolution methods, such as Richardson-Lucy,
do not provide satisfactory results. 
We have therefore proposed and analyzed a new method of image reconstruction, which we call 
multi-component Richardson-Lucy 
(MC-RL), in which we consider the image to be reconstructed as the sum of two terms (star plus jet). 
A regularization parameter is introduced for this second component and the 
reconstruction is obtained with an iterative method alternating between the two 
components. 

Our main conclusions can be summarized as follows:

\begin{itemize}
\item The proposed MC-RL method works better than the standard RL
method since it is able to effectively reduce artifacts in the star region, which is the
most important one in the light of the scientific aim of the simulated observations, that
is the study of the formation and collimation of the jet.
\item For both methods the computational cost per iteration is high since one has to
manage four 1024$\times$1024 LN-images. Therefore it is important to increase
the efficiency of the algorithm by reducing the number of iterations. To
this purpose the SGP approach proposed by Bonettini et al. {\cite{BZZ}} seems
to be quite promising.
\item The choice of a suitable value of the regularization parameter $\mu$ is
an important issue. In this paper the parameter is estimated by searching for
a minimum of the r.m.s. error. The value we found is also providing the best results
for what concerns the measurements of both the flux and width of the jet, 
as confirmed by our comparison of the original and reconstructed images. 
We therefore show that the value of $\mu$ can be successfully fine-tuned in our simulations,
although the problem of estimating the optimal value of $\mu$ might not be trivial 
in the case of real data.
\item Besides this problem it is necessary to evaluate the robustness of the
method with respect to errors in the estimate of the position of the star and
evaluate its accuracy as a function of the relative magnitude of the star and
the surrounding jet.
\item Finally, the method can be easily extended to the case of several 
stars superimposed to smoothly varying objects, provided the positions of the
stars can be estimated with sufficient accuracy.
\item As a future development of this work, we plan to investigate the results provided by
the MC-RL method when the contrast between star and jet is even higher, 
which is often the case for real objects in the sky.
\end{itemize}

%%%%%%%%%%%%%%%%%%%%%%%%%%%%%%%%%%%%%%%%%%%%%%%%%%%%%%%%%%%%%
\acknowledgments     %>>>> equivalent to \section*{ACKNOWLEDGMENTS}       

This work has been partially supported by MIUR (Italian Ministry for University 
and Research), PRIN2008 ``Optimization Methods and Software for Inverse 
Problems'', grant 2008T5KA4L, and by INAF (National Institute for Astrophysics) 
under the contract TECNO-INAF 2010 ``Exploiting the adaptive power: a dedicated 
free software to optimize and maximize the scientific output of images from 
present and future adaptive optics facilities''.

%%%%%%%%%%%%%%%%%%%%%%%%%%%%%%%%%%%%%%%%%%%%%%%%%%%%%%%%%%%%%
%%%%% References %%%%%

\bibliographystyle{spiebib}   %>>>> makes bibtex use spiebib.bst
\bibliography{spie2012_yso}   %>>>> bibliography data in report.bib

\end{document}